\documentclass[10pt,twocolumn]{paper}
\usepackage{epsfig, graphicx}
\usepackage[english]{babel}
\usepackage[margin=1.5cm]{geometry}

\title{\center \rm \bf  Prospects for the Use of Laboratory Sources for X-ray Reflectometry of Thin Films on a Liquid Surface}

\author{\small \rm Viktor E. Asadchikov$^{a}$\/\thanks{acad@crys.ras.ru}, Yuri O. Volkov$^{a}$\/\thanks{neko.crys@gmail.ru}, Alexander D. Nuzhdin$^{a}$, Boris S. Roshchin$^{a}$ and Aleksey M. Tikhonov$^{b}$\/\thanks{tikhonov@kapitza.ras.ru}\\
 \small $^a$ Shubnikov Institute of Crystallography, Federal Research Center Crystallography and Photonics,
Russian Academy of Sciences, \\
\small Moscow, 119333 Russia\\
\small $^b$Kapitza Institute for Physical Problems, Russian Academy of Sciences, Moscow, 119334 Russia\\
}

\begin{document}
\maketitle

\abstract{ \it  The authors present a review of the systematic studies of the structure of macroscopically planar thin films at the air-liquid interface (water, alkali solution and silica hydrosol). A common feature of the considered works is the application of a model-independent approach to the analysis of X-ray reflectometry data, which does not require a priori assumptions about the structure of the object under study. It is shown that the experimental results obtained with the laboratory source in some cases are qualitatively on par with the results of those obtained with the use of synchrotron radiation source. The reproducibility of the effect of spontaneous ordering in films of amphiphilic organic molecules (phospholipids) at the surface of the colloidal solution of silica nanoparticles is demonstrated. The possibility of influencing the kinetics of the in situ formation of a phospholipid film by enriching the liquid substrate with alkali metal ions is also discussed. }
\vspace{0.25in}

{\bf 1. INTRODUCTION}

Phospholipids at the water surface form a planar film – Langmuir monolayer – which is the simplest model of a cell membrane for biophysical studies \cite{1}. The main method traditionally used for studying the structure of such systems is X-ray reflectometry and scattering under grazing incidence conditions, due to its non-destructive nature and high sensitivity to surface effects. A significant number of publications on the studies of Langmuir lipid monolayers on the surface of the water by X-ray methods is presented in
literature, for example, in works \cite{2,3}; a more recent review of the topic can be found, for example, in \cite{4}.

However, studies of such samples are limited by a number of features. First of all, contrast -- the relation between densities of lipid mesophase and aqueous substrate -- in X-ray experiments is in the range of $0.95 \div 1.05$ \cite{5}, which leads to
the necessity for high beam intensity and low experimental error of the measured signal. As a result, reflectivity and scattering experiments from Langmuir systems are carried out at specialized synchrotron stations. Further, the need for a horizontal orientation of the sample
under investigation imposes restrictions on the design of optical path for the x-ray station. At the same time, for a synchrotron scource, a small illuminated area on a sample combined with high radiation intensity leads to degradation of the film under probing beam in a time interval comparable to that of a single measurement \cite{6}. 

A separate problem is the preparation of samples of more complex lipid structures, especially in the form of bilayers and multilayers. Indeed, the characteristic radius of the spontaneous curvature of a lipid bilayer in an aqueous environment is about $10$~$\mu$m, which leads to the formation of macroscopic three dimensional aggregates (liposomes and vesicles). As a result, the works on structural analysis of multilayer lipid membranes are limited to a samples of vesicular lamellae on solid substrates \cite{7,8}. At the same time, it has been previously reported that ordered lipid lamellar films can form on the highly polarized surface of aqueous solutions of amorphous silica \cite{9}.

Here we present a review of our systematic studies of macroscopic lipid films at the surface of silica hydrosol substrates by X-ray
reflectometry. A key feature of these works is the usage of laboratory X-ray diffractometer with movable emitter-detector system and horizontal sample location \cite{10} to perform experimental measurements of reflectivity. The possibility of mutually independent movement of both the source and the detector in relation to a stationary sample allowed us to significantly simplify the design of optical system. The second significant feature is the application of the model-independent approach to the processing and analysis of X-ray reflectometry and scattering data \cite{11,12}.
\vspace{0.25in}

{\bf 2. EXPERIMENT DESCRIPTION AND DATA PROCESSING}

Liquid substrates were prepared in a fluoroplastic dish with diameter of 100~mm installed in a sealed cell with X-ray transparent windows, in accordance with the procedure described in \cite{9}. A fixed volume of a solution of phospholipid in chloroform ($\sim 50$\,mmol/l) was applied on the surface of the substrate by the droplet method using a calibrated Hamilton syringe. The required volume of the solution ($\sim 10$~$\mu$l) was calculated for the amount of substance, when being completely spreaded over the surface, to be sufficient for the formation of more than 10 monolayers of lipid. A change in the surface tension $\gamma$ from 74~mN/m to $\sim 50$~mN/m, accompanying the process of droplet spreading, was recorded by the Wilhelmy method (NIMA PS-2). After preparation, each
sample was kept at room temperature ($T \approx 298$~K) for at least an hour to bring it to a thermodynamic equilibrium. 

The design of a general-purpose laboratory diffractometer with movable emitter-detector system (DRS) is described in detail in \cite{10}. A wide-focus ($12\times2$~mm) X-ray tube with a copper anode was used as a radiation source. The probe radiation was prepared by a single-reflection crystal monochromator Si(111), tuned to the $K_{\alpha1}$ line of copper (photon energy $E \approx 8048$\,eV, wavelength $\lambda = 1.5405 \pm 0.0001$\,{\AA}), and a vacuum triple-slit collimation system, which allowed us to achieve beam linear width (intensity distribution in the plane of specular reflection) $d \approx 0.55$\,mm with an integral beam intensity of
$3\times10^6$\,counts/s. The scintillation detector Radicon SCSD-4 (with noise level 0.1 counts/s) was used to register the signal. Thus, the measurement range for the decrease in the signal intensity $R_{max}/R_{min}$ was 7 to 8 orders of magnitude, which is
comparable with measurements at 2nd-generation synchrotron sources. 

\begin{figure}
\hspace{0.15in}
\epsfig{file=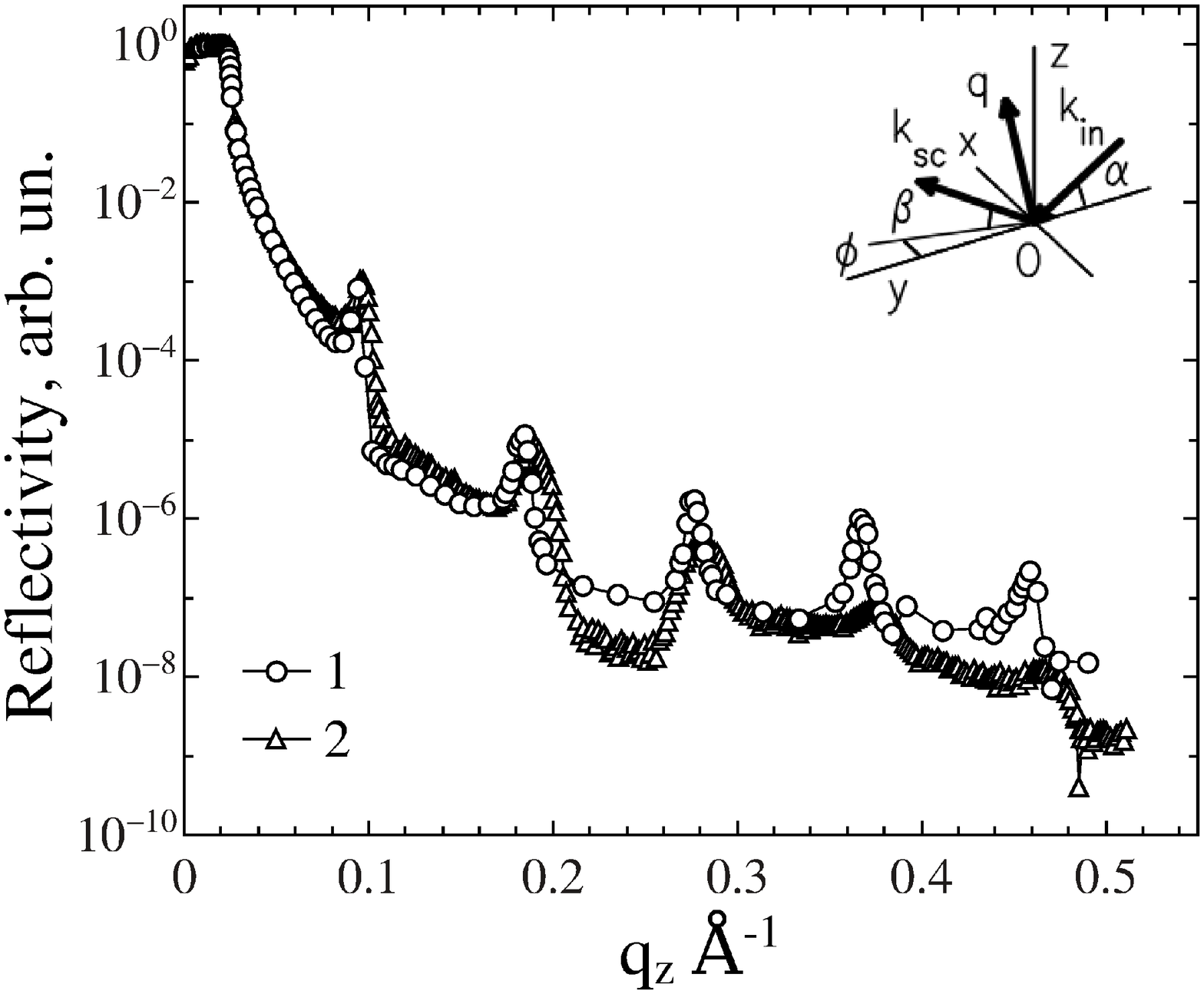, width=0.45\textwidth}

\small {\bf Figure 1.} \it X-ray reflectivity $R(q_z)$ for planar DSPC multilayer at the silica hydrosol surface measured with the use of the DRS diffractometer (1) and at the X19C beamline of the NSLS synchrotron (2). The data taken from \cite{21,22}. Insert: kinematics of X-ray scattering at the air -- liquid  interface.

\end{figure}
\vspace{0.25in}

Fig. 1 shows an example of the experimental angular dependencies of specular reflectivity $R(q_z)$ obtained on a DRS
diffractometer (Fig. 1, curve 1) and at the X19C beamline ($E \approx 15$\,keV) of the National Synchrotron Radiation Scource, Brookhaven  National Laboratory, USA (Fig. 1, curve 2). The methods of the measurements and the processing of the obtained data is described in more detail, for example, in \cite{13}.

 Kinematics of X-ray scattering at a macroscopically flat horizontal surface under grazing incidence conditions can be conveniently described in the coordinate system where the center of illuminated area corresponds to the origin point $O$, the $xy$ plane coincides with the air-sample interface, and the $Oz$ axis is normal to the surface (see the inset on Fig. 1). Here {\bf k}$_{in}$ and {\bf k}$_{sc}$ are the wave vectors of the incident and scattered rays, $\alpha$ and $\beta$ are the grazing and scattering angles ($\alpha,\beta \ll 1$), $\phi$ is the angle of azimuthal deviation of the scattered ray. Under conditions of specular reflection ($\alpha=\beta$ and $\phi = 0$), the scattering vector has a single component $q_z =|${\bf k}$_{in}$ - {\bf k}$_{sc}|$ $= (4\pi/\lambda)\sin\alpha$. The angular dependence of the specular reflection factor, in its turn, has the form  $R(q_z)=R_F(q_z)|\Phi(q_z)|^2$, where $R_F$ is the reflectivity from the ideal air-substrate interface, and $\Phi(q)=\frac{1}{\Delta\rho}\int^{+\infty}_{-\infty}\left\langle\frac{d\rho(z)}{dz}\right\rangle e^{iqz} dz$  is the structural factor for the distribution of electron density $\rho(z)$ depth-wise along the $Oz$ axis, averaged over the illumination area.
 
For the analysis of the experimental specular reflectivity data $R(q_z)$ and reconstruction of the electron density distribution $\rho(z)$ we applied a model-independent approach proposed in \cite{11}, which is based on the extrapolation of the asymptotic component of reflectivity $R$ into the range of large angles (where $q_z > q_{max}$). Unlike classical approaches which are based on the
parametrical optimization of a theoretical model of the object, the model-independent approach does not require any a priori assumptions about the structure under investigation, and allows us to directly calculate the permittivity distribution $\varepsilon(z)$ (and, respectively, the volumetric electron concentration $\rho(z) = \pi(1- \varepsilon(z))/(r_0\lambda^2)$, where $r_0$ is the classical electron radius) depth-wise in the direction normal to the interface plane. Features of the approach, the problem of uniqueness of the solution of inverse problem for the reflectometry case, and the calculation algorithm are described in detail in \cite{12}.
\vspace{0.25in}

{\bf 3. STRUCTURAL EFFECTS ON THE SURFACE OF SILICA HYDROSOL}

The structure of the transition layer on the surface of a silica hydrosol -- a colloidal solution of SiO$_2$ nanoparticles in water stabilized by alkali (NaOH) -- has been previously investigated by one of the authors in \cite{14, 15} only in frames of an analytical model. The air -- sol interface in that kind of a system is strongly polarized in the direction normal to the surface, due to the difference in potentials of the "electric image" effect between macro-nanoparticles that carry large charge ($\sim 10^3$ electrons) and alkaline Na$^+$ ions. This leads to separation of the planes of closest approach for ions and nanoparticles.

 \begin{figure}
\epsfig{file=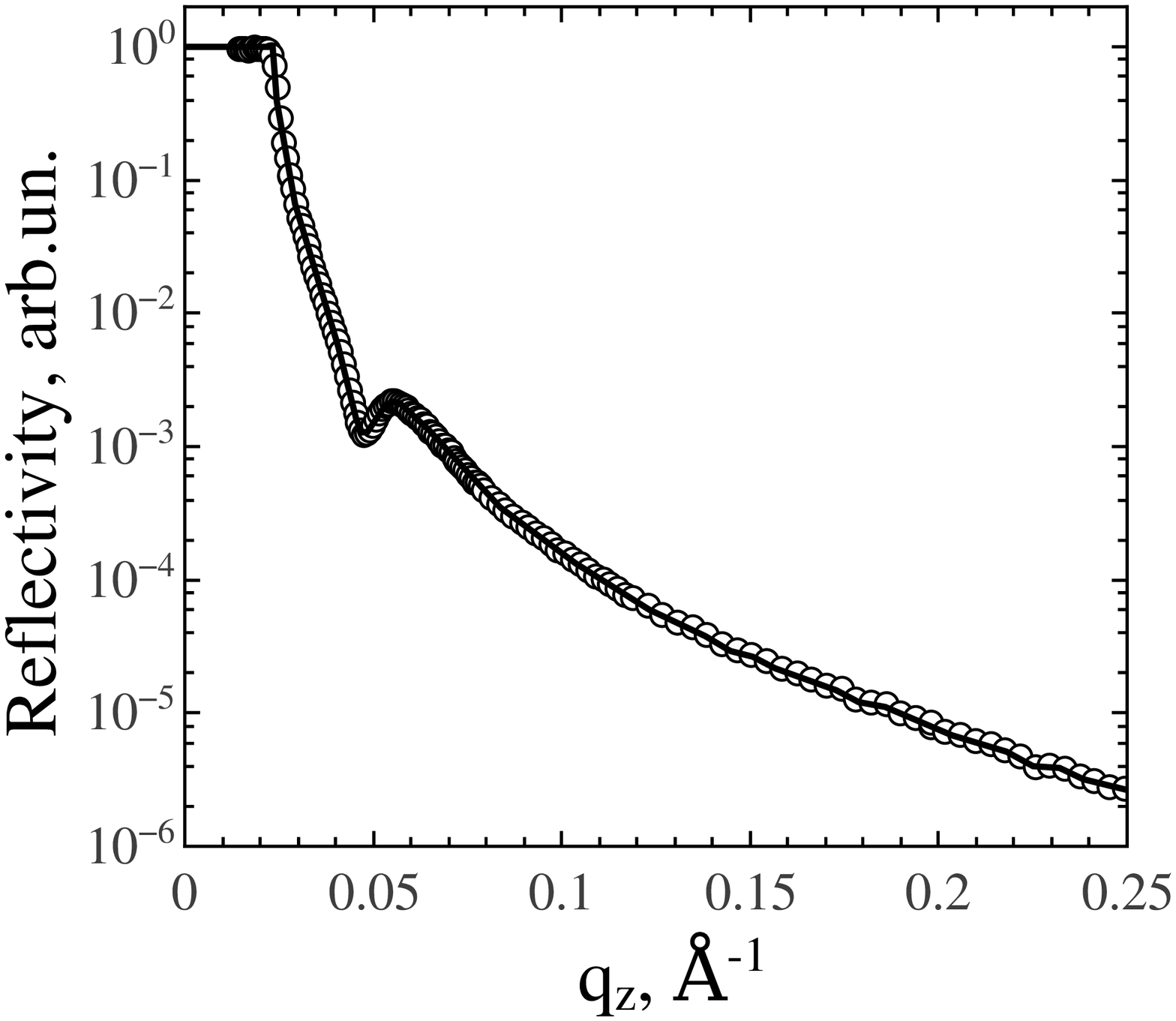, width=0.25\textwidth}
\epsfig{file=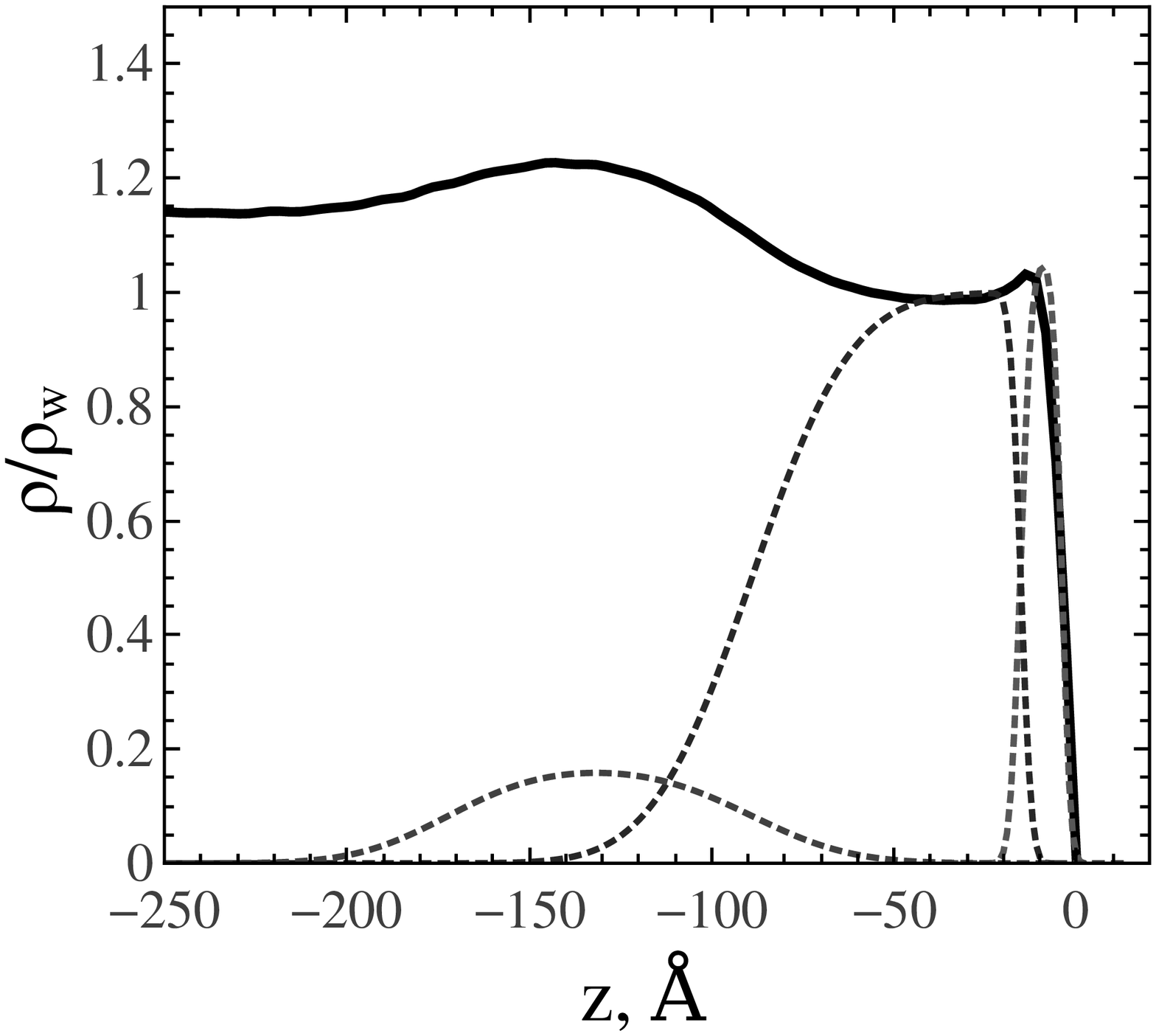, width=0.25\textwidth}

\small {\bf Figure 2.} \it a(left): Reflectivity curve $R(q_z)$ from the SM-30 silica sol surface.
b(right): Normalized profile $\rho(z)/\rho_w$  (solid line) and model decomposition (dashed line). The data taken from \cite{16}.

\end{figure}
\vspace{0.25in}
  
In \cite{16} we examined the structure of surface layering for colloidal solutions of Ludox SM-30 (30\% wt. SiO$_2$ and 0.2\% wt. Na$^+$) and TM-50 (50\% wt. SiO$_2$ and 0.3\% wt. Na$^+$), as well as the effect of rearrangement of the structure after
the application of a model lipid 1,2-distearoyl-sn-glycero-3-phosphocholine (DSPC) on it. Note that the characteristic diameter of the silicon oxide particles, previously calculated from the small-angle X-ray scattering data, was $\sim 12$\,nm for
the SM-30 solution and $\sim 27$\,nm for the TM-50 solution. An example of X-ray reflectivity curve $R(q_z)$, measured from the SM-30 solution, is shown in Fig. 2a (circles are the experimental points, solid line illustrates accuracy of the reconstruction).
The model-independent distribution of electron concentration $\rho(z)$ , calculated and normalized to the volumetric electron concentration in deionized water $\rho_w=0.333$ {\it e$^-$/}{\AA}$^3$ (solid line in Fig. 2b) is in good agreement with the analytical model from \cite{14}: a "suspended"{} Na$^+$ ions on the surface, a depleted layer of water, and an "macroion wall"{} of SiO$_2$ nanoparticles (dashed lines in Fig. 2b is a water content in this layer).

\begin{figure}
\epsfig{file=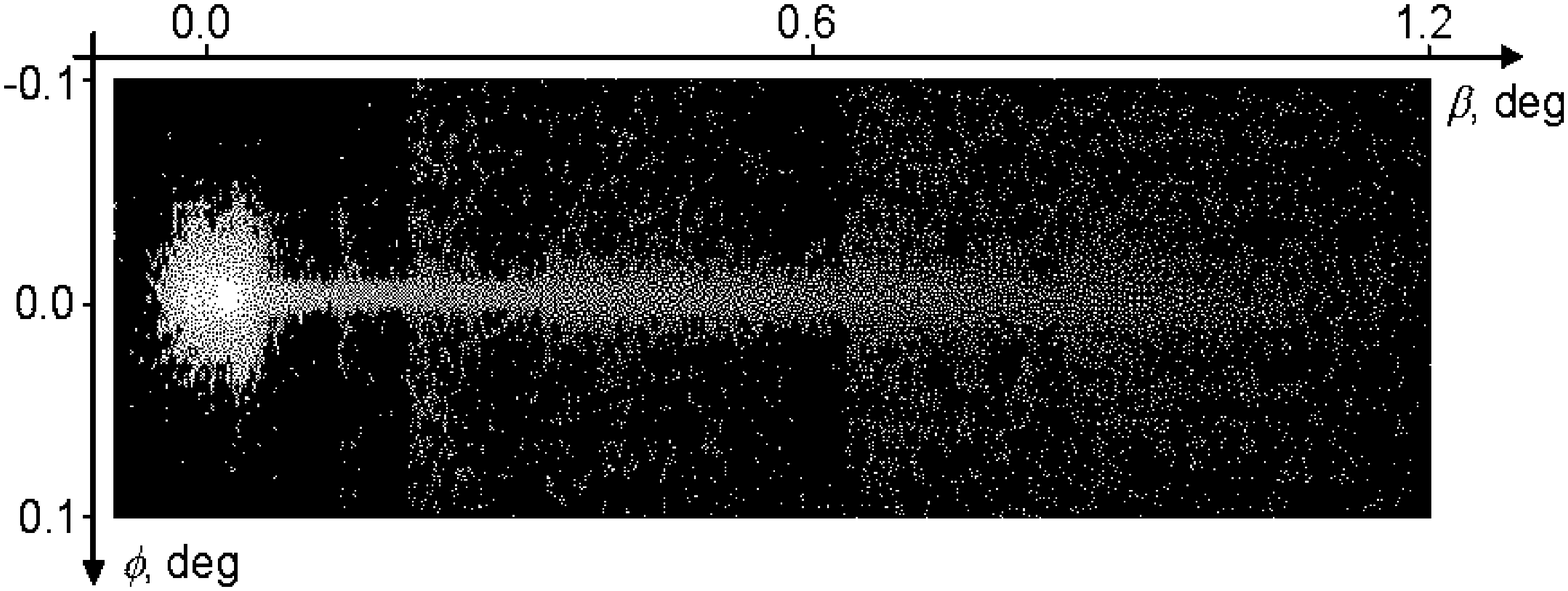, width=0.5\textwidth}
\hspace{0.4in}
\epsfig{file=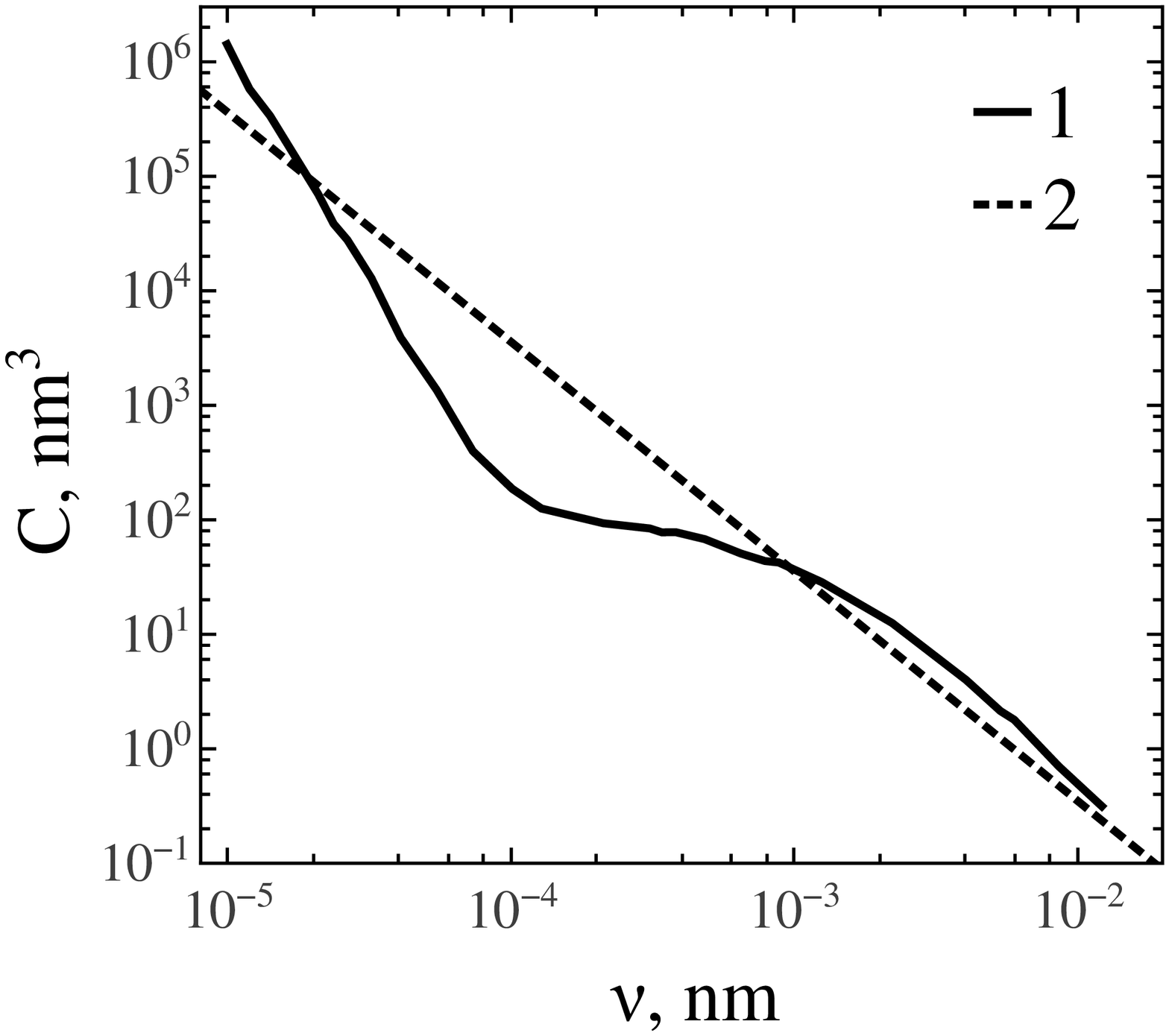, width=0.4\textwidth}

\small {\bf Figure 3.} \it a) Two-dimensional distribution of diffuse scattering from the surface of TM-50 silica sol.
b(bottom): Calculated function $C(\nu)$ (1) ($\nu$ is the spatial frequency) and the theoretical estimate in the frames of capillary roughness model $C_{cap}(\nu)$ (2). The data taken from \cite{17}.
\end{figure}
\vspace{0.25in}

Note that the roughness of the interface can have a significant effect on the dynamics of surface processes. In \cite{17}, we carried out a comprehensive study of the surface structure of TM-50 silica sol accounting for surface roughness by the diffuse scattering. It was found that the statistical distribution function of roughness heights $C(\nu)$ (the so-called "power spectral density function"{}  as a function of  the spatial frequency, $\nu$, \cite{18}) calculated from the angular distribution of scattered radiation, differs significantly from the theoretical predictions of the capillary waves theory, which is widely used in the literature \cite{19} (Fig. 3). We assume that this effect is due to the influence of viscosity of colloidal solution in the near-surface region. This assumption is discussed in more detail in \cite{20}, where we carried out observations of the whispering gallery effect at water and silica sol surfaces, and also analyzed the dynamics of the efficiency of intensity transfer along the liquid meniscus depending on the surface roughness \cite{WG1,WG4}.
\vspace{0.25in}

{\bf 4. MULTILAYERS OF PHOSPHOLIPIDS ON A LIQUID SURFACE}

Kinetics of the spontaneous ordering of phospholipid multilayers on the surface of silica sol was discussed in our works \cite{21,22}. We
used model lipids 1,2-distearoyl-sn-glycero-3-phosphocholine (DSPC) and 1-stearoyl-2-oleoylsn-glycero-3-phosphocholine (SOPC) deposited on the surface of silica sols FM-16 (16\% wt. SiO$_2$ and 0.2\% wt. Na$^+$, characteristic diameter of nanoparticles 5\,nm) and SM-30 (see above). Note that DSPC and SOPC lipids have different temperatures $T_c$ of a phase transition associated
with chain melting \cite{1}: at room temperature (295\,K) DSPC is in the gel phase, and SOPC is in the melted phase. When the DSPC film has been kept in thermodynamic equilibrium for about 24 hours, a regular set of diffraction peaks with a period $\Delta q_z
= 2\pi/d$ appears on the angular dependence of the reflection factor (curve 1 in Fig. 4a), where the characteristic thickness $d$ of a structural element in the multilayer corresponds to the thickness of DSPC bilayer in the crystalline phase $L \approx 68$\,{\AA}
known in literature.

\begin{figure}
\epsfig{file=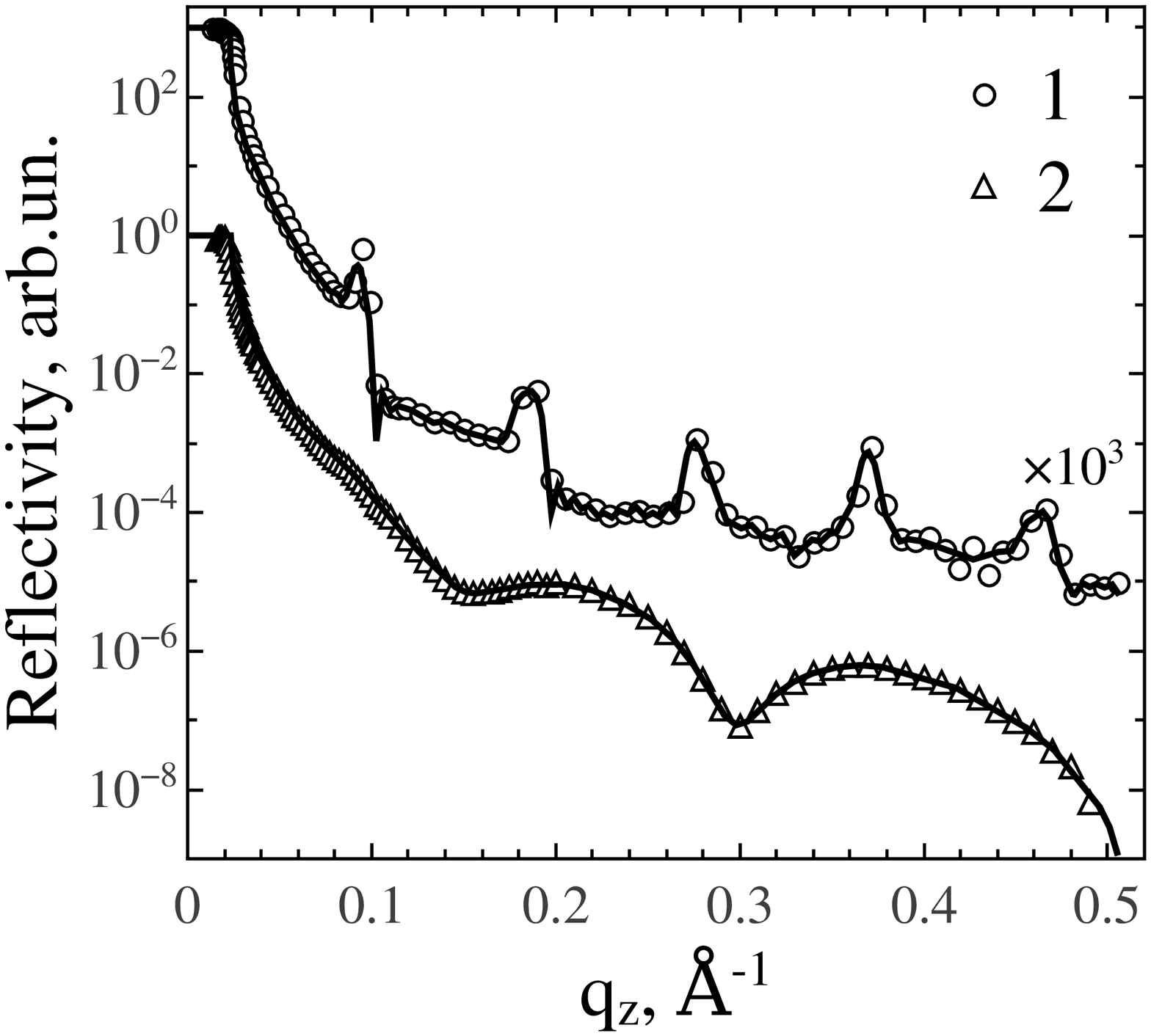, width=0.25\textwidth}
\epsfig{file=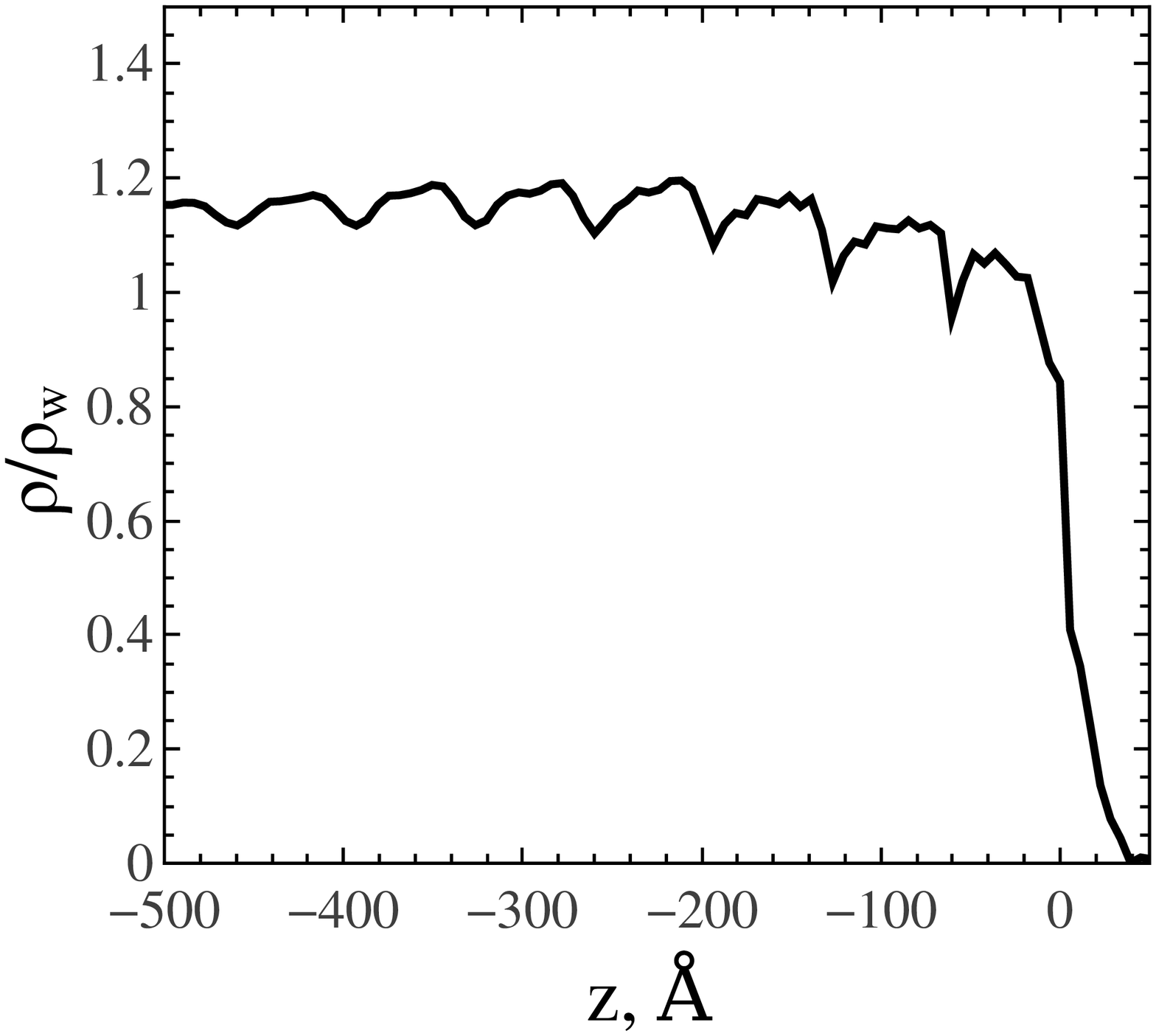, width=0.25\textwidth}

\hspace{0.4in}
\epsfig{file=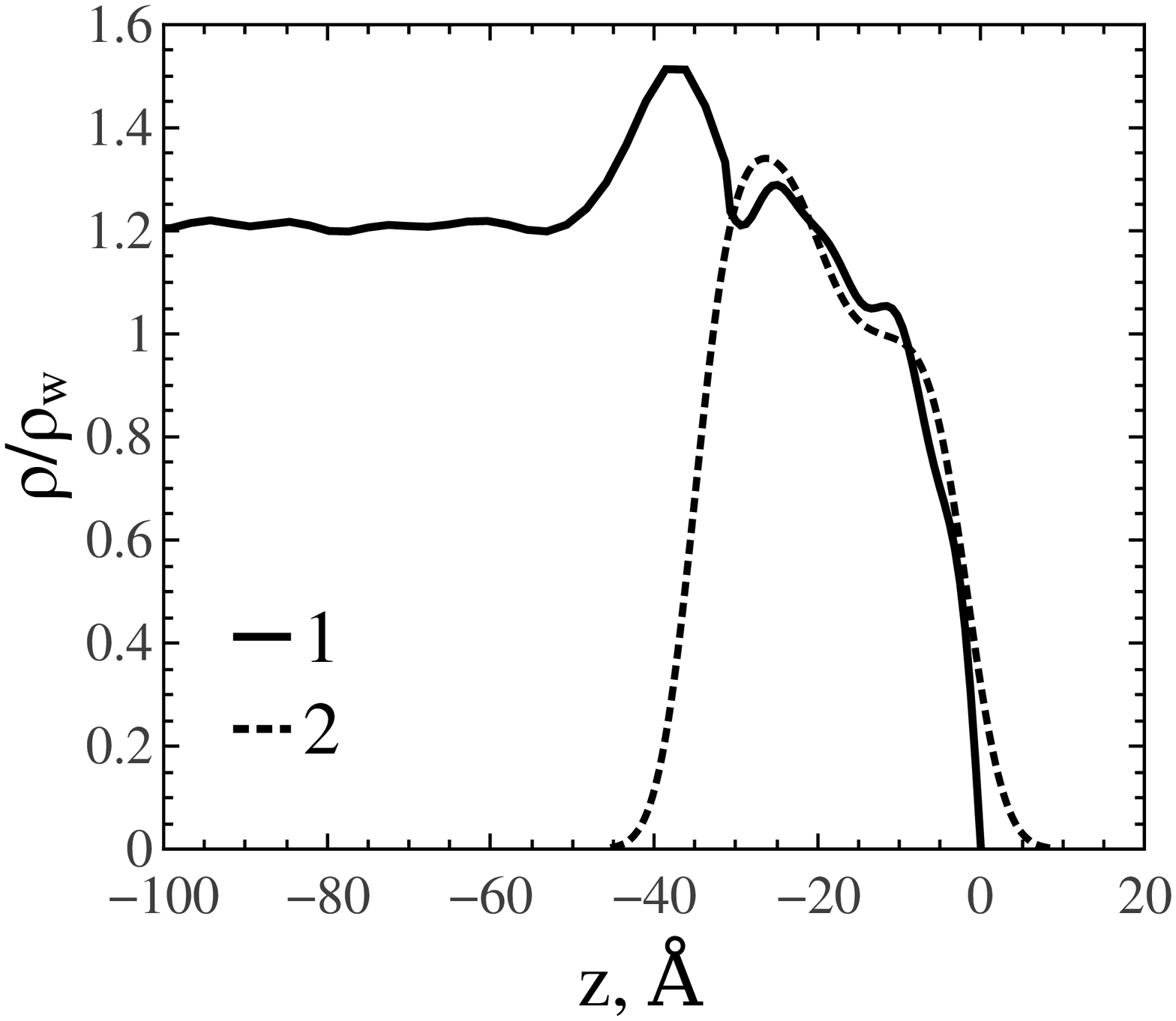, width=0.4\textwidth}

\small {\bf Figure 4.} \it a(top left): Reflectivity curves $R(q_z)$ from DSPC film at the surface of original SM-30 silica sol (1) and one enriched with NaOH (2). b(top right): Normalized profile$\rho(z)/\rho_w$ for DSPC film at the silica sol SM-30. Insert: lipid multilayer structure. c(bottom): Normalized profile $\rho(z)/\rho_w$ for DSPC film at the silica sol surface enriched with NaOH (1) and theoretical profile for the Langmuir monolayer of DSPC (2). The data taken from \cite{21,22}.

\end{figure}
\vspace{0.25in}

According to the reconstructed electron concentration profile $\rho(z)$ normalized to the electron concentration in deionized water $\rho_w=0.333$ {\it e$^-$/}{\AA}$^3$ (Fig. 4b), the total thickness of the structure agrees well with the Debye screening
length in the bulk hydrosol $\Lambda_D \approx 500$\,{\AA}. At the same time, the electron concentration in each of the periodic layers exceeds the theoretical value for density distribution along the lipid molecule. A number of authors previously assumed, according to the results of molecular dynamics modeling, that an additional potential in the lipid membrane can arise due to the penetration of Na$^+$ ions into it from the bulk of substrate \cite{23}. Assuming that DSPC molecules in the bilayers are in three-dimensional crystalline phase with the most dense packing corresponding to the area per molecule $A = 41.6$\,{\AA}$^2$, the calculated value of excess electron concentration in the multilayer corresponds to $\sim 9$ Na$^+$ ions per each lipid molecule. However, according to
the formation time of the ordered structure, the estimated resistivity of the multilayer per area unit is lower by 4-5 orders of magnitude
than the values known in the literature from measurements of the ionic conductivity of lipid films on water and solid substrates. We assumed in \cite{22} that a more efficient transfer of ions from the volume of silica sol to the multilayer is caused by electroporation of the lipid film under the influence of an electric field on the sol surface, which significantly exceeds the theoretical limit for its electrical stability.

The influence of chemical composition of a silica substrate on the structure and properties of the lipid membrane has been also considered in \cite{21}. In particular, the enrichment of SM-30 substrate with alkaline ions (up to 1.3\% wt. NaOH) before applying a lipid film onto it leads to the disappearance of diffraction peaks on the specular reflectivity $R(q_z)$ (curve 2 in Fig. 4a), which corresponds to the collapse of a lipid structure (to the monolayer state) and to condensation of nanoparticles on it from the bulk (profile 1 in Fig. 4c). In this case, excess lipid on the surface forms macroscopic bulk aggregates that persist in equilibrium with the lipid film and are condensed on the edge of meniscus. The decrease in thickness of the lipid layer is also consistent with the calculated decrease in the Debye screening length $\Lambda_D$  (up to $\sim 100$\,{\AA}) in an alkaline-enriched solution. The estimated area per lipid molecule $A$ calculated from the integrated electron density is $45 \pm 2$\,{\AA}$^2$, which corresponds to the value for Langmuir monolayers in two-dimensional
liquid crystal phase \cite{9,24}. Thus, the possibility of controlling the thickness and phase state of a lipid film by changing the concentration of alkaline ions in a liquid substrate was shown in \cite{21}.

It should be noted that the films of saturated phospholipids exhibit a first-order phase transition between the states of "liquid expanded"{} (LE) and "two-dimensional liquid crystal"{} (LC) under changes in surface conditions, in particular, lateral pressure \cite{24}. As a result, for the correct interpretation of the structure of mono and bilayers, it is necessary to apply various research methods, including those providing a qualitative model of the structure. In \cite{25, 26} we investigated the LE-LC phase transition in the 1,2-dimyristoyl-sn-glycero-3-phospho-L-serine monolayer (DMPS) on the surface of a KCl solution in deionized water using an
integrated approach which included methods of X-ray reflectometry (XR) and molecular dynamics simulations (MD).

As an example Fig. 5a shows the angular dependence of the reflection factor for a compressed DMPS monolayer in the liquid crystal phase with an estimated area per molecule $A\approx 45 $\,{\AA}$^2$. Fig. 5b shows the electron density profile $\rho(z)$ calculated by the model-independent approach (solid line) and the decomposition profiles of the structural elements from a theoretical monolayer
model (dashed lines). In the region of molecular lipid "tails"{} adjacent directly to the surface, the distribution $\rho(z)$ corresponds with good accuracy to the highly ordered structure of hydrocarbon chains with an angle of deviation from the normal to the surface $\theta \approx 26$\,deg. However, in the region of the polar groups of phosphatidylserine (peak of $\rho(z)$ in Fig. 5b) the integral electron concentration for model-independent calculation exceeds notably the theoretical value. This effect is supposedly caused by hydration of the polar groups; the calculation of the excess number of electrons corresponds to $\sim 5$ H$_2$O molecules per
each lipid molecule. Note that this estimate almost coincides with the modeling of water distribution in the structure of lipid membrane according to MD calculations. In \cite{26}, the effects of hydration of a lipid film for various values of area per molecule $A$ during compression are discussed in more detail.
\vspace{0.25in}

{\bf 5. CONCLUSION}

Thus, in a series of publications \cite{13,16,17,21,22,25,26} we systematically demonstrated the possibilities of studying the
structure of macroscopically flat phospholipid films at the air-liquid interfaces by X-ray reflectometry performed on a laboratory source.
A key feature of these works is the analysis of experimental data within the frames of the model-independent approach, which allows us
to obtain information on the transverse structure of films directly without involving any a priori models. The formation of a multilayer lipid membrane on the surface of colloidal solutions (silica sol) was reproduced repeatedly; the timewise dynamics of the spontaneous ordering of multilayer, as well as the possibility of influencing the electrical properties of air-silica sol interface and the structure of the formed membrane by enriching the substrate with alkali metal ions, are investigated. For the first time we demonstrated the whispering gallery effect at the liquid samples. In addition to, we considered the deviation of the experimental statistical parameters of the surface
roughness from the theoretical predictions of the standard theory of capillary waves.

\begin{figure}
\epsfig{file=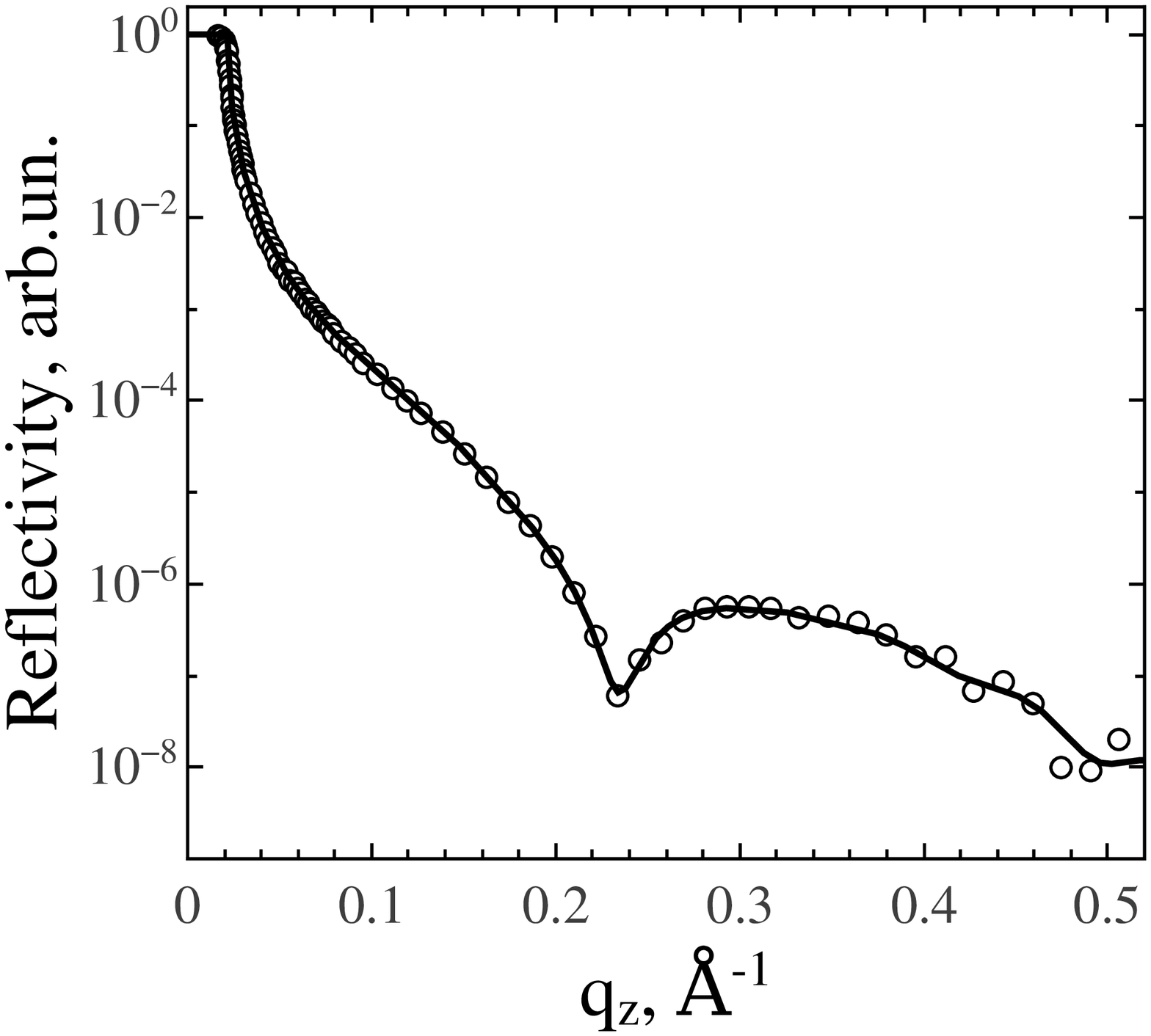, width=0.25\textwidth}
\epsfig{file=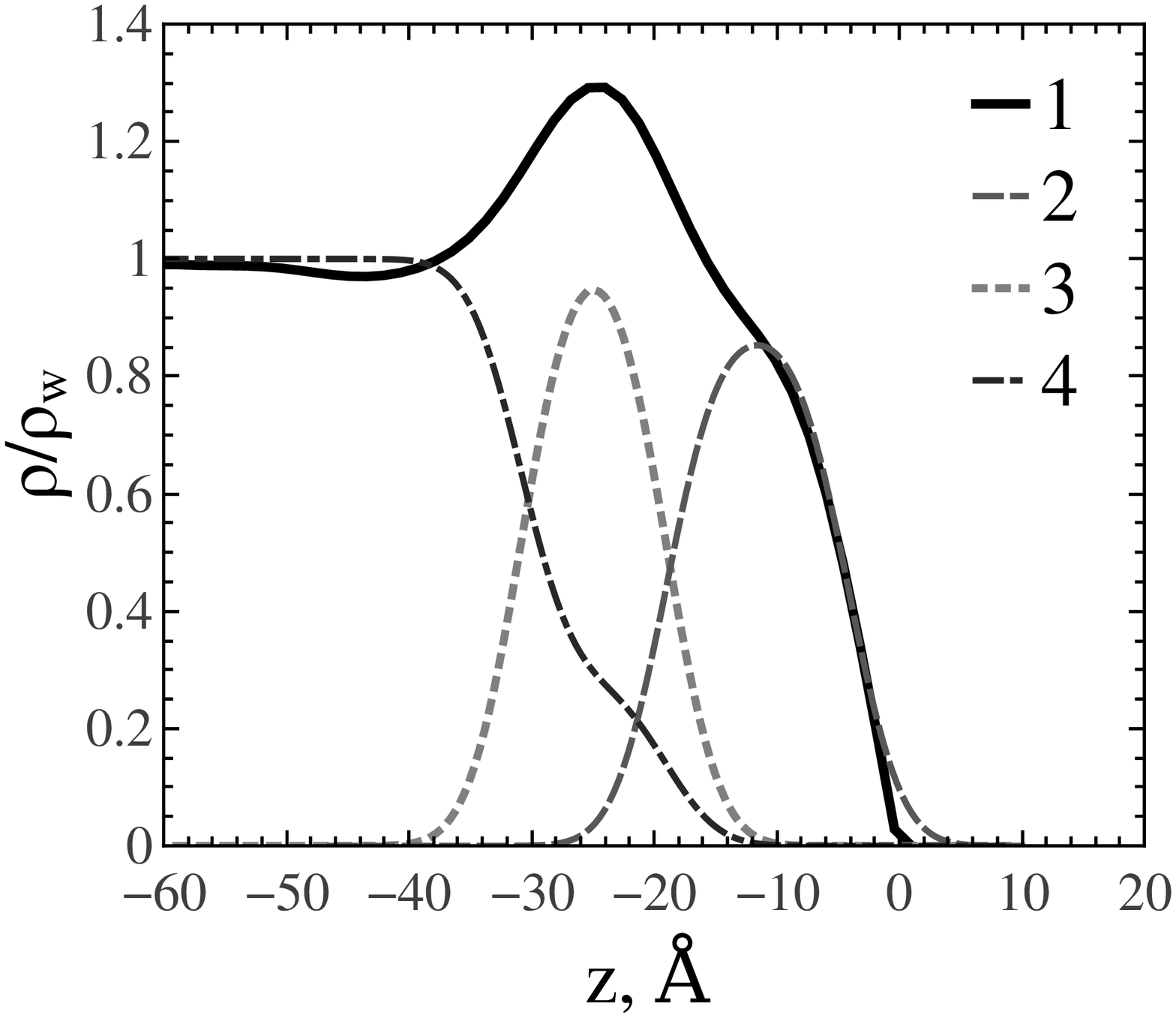, width=0.25\textwidth}

\small {\bf Figure 5.} \it a(left): Reflectivity curve $R(q_z)$ from the DMPS monolayer in the LC phase on the surface of the water. b(right): Normalized profile $\rho(z)/\rho_w$ (1) and decomposition of the theoretical model of MD: a layer of hydrocarbon "tails" (2), a layer of lipid polar groups (3), water (4). The data taken from \cite{26}.

\end{figure}
\vspace{0.25in}

As it is shown, the application of the model-independent method for reconstruction of the structure can serve as an independent confirmation of the correctness of the mathematical modeling of such films, including molecular dynamics simulations. With a comparable range of both the ratio of the incident beam to the measured signal intensity ($R_{max}/R_{min} \sim 10^8$) and scattering
vectors ($q_{max} \sim 0.5$\,\AA$^{-1}$), the non-destructive nature of laboratory measurements using reflectometry is a promising method for studying the in situ structure of organic films on the surface of liquid substrates. We also believe that it is useful to apply our
approach in combination with other experimental and theoretical methods simultaneously, for example, to study the processes of adsorption of macromolecules (proteins or polymers) on a phospholipid monolayer, which can be an important step in understanding the mechanisms of functioning of biological membranes.
\vspace{0.25in}

{\bf 5. ACKNOWLEDGEMENTS}

The authors are grateful to Yu.\,A.\,Ermakov and A.\,M.\,Nesterenko for fruitful discussions, useful comments, and help with the experiments. The work has been supported by the Ministry of Science and Higher Education of the Russian Federation within the State assignment of FSRC "Crystallography and Photonics" RAS and IPP RAS.


\begin{thebibliography}{49}
\small

\bibitem{1}
D.\,M.\,Small, {\it The Physical Chemistry of Lipids}, Plenum Press, New York, 1986.

\bibitem{2}
H.\,Mohwald, {\it Phospholipid and phospholipid-protein monolayers at the air/water interface}, Annu. Rev. Phys. Chem. 41, 441 (1990).

\bibitem{3}
V.~M.~Kaganer, H.~Mohwald, and P.~Dutta, {\it Structure and phase transitions in Langmuir monolayers}, Rev. Mod. Phys. 71, 779 (1999).

\bibitem{4}
M.~Delcea M and C.~A.~Helm, {\it X-ray and neutron reflectometry of thin films at liquid interfaces}, Langmuir 35(26), 8519 (2019).

\bibitem{5}
N.~Kucerka, J.~Nagle , J.~N.~Sachs, S.~E.~Feller, J.~Pencer, A.~Jackson and J.~Katsaras, {\it Lipid bilayer structure determined by the simultaneous analysis of neutron and X-ray scattering data}, Biophys. J. 95, 2356 (2008).

\bibitem{6}
S.~M.~Danauskas, M.~K.~Ratajczak, Yu.~Ishitsuka, J.~Gebhardt, D.~Schultz, M.~Meron, K.~Yee, C.~Leea and B.~Lin,
{\it Monitoring X-ray beam damage on lipid films by an integrated Brewster angle microscope/X-ray diffractometer}, 
Rev. Sci. Instr. 78, 103705 (2007).

\bibitem{7}
J.~Nagle, R.~Zhang, S.~Tristram-Nagle, W.~Sun, H.~I.~Petrache and R.~M.~Suter, {\it X-ray structure determination of fully hydrated L-alpha phase dipalmitoylphosphatidylcholine bilayers}, Biophys. J. 70, 1419 (1996).

\bibitem{8}
G.~Forster, C.~Schwieger, F.~Faber, T.~Weber  and A.~Blume,  {\it Influence of poly(L-lysine) on the structure of dipalmitoylphosphatidylglycerol/water dispersions studied by X-ray scattering}, Eur. Biophys. J. 36, 425 (2007).

\bibitem{9}
A.\,M.\,Tikhonov, {\it Multilayer of phospholipid membranes on a hydrosol substrate}, JETP Lett. 92, 356 (2010); arXiv:1010.1680 [cond-mat.soft].

\bibitem{10}
V.\,E.\,Asadchikov, V.\,G.\,Babak, A.\,V.\,Buzmakov, Yu.\, P.\, Dorokhin, I.\, P.\, Glagolev, Yu.\, V. \,Zanevskii,
V.\,N.\,Zryuev, Yu.\,S.\,Krivonosov, V.\,F.\,Mamich, L.\,A.\,Moseiko, N.\,I.\,Moseiko, B.\,V.\,Mchedlishvili, S.\,V.\,Savel'ev, R.\,A.\,Senin, L.\,P.\,Smykov, G.~A.~Tudosi, V.~D.~Fateev, S.~P.~Chernenko, G.~A.~Cheremukhina, E.~A.~Cheremukhin,  A.~I.~Chulichkov, Y.~N.~Shilin, and V.~A.~Shishkov, {\it An X-ray diffractometer with a mobile emitter-detector system},
Instrum. Exp. Tech. 48, 364 (2005).

\bibitem{11}
I.~V.~Kozhevnikov, {\it Physical analysis of the inverse problem of X-ray reflectometry} Nucl. Instrum. Methods Phys. Res., Sect. A 508, 519 (2003).

\bibitem{12}
 I.\,V.\,Kozhevnikov, L.\,Peverini, E.\,Ziegler, {\it Development of a self-consistent free-form approach for studying the three-dimensional morphology of a thin film}, Phys. Rev. B 85, 125439 (2012).

\bibitem{13}
V.~E.~Asadchikov,  A.~M.~Tikhonov,  Yu.~O.~Volkov, B.~S.`Roshchin, Yu.~A.~Ermakov, E.~B.~Rudakova,
I.~G.~D'yachkova and A.~D.~Nuzhdin, {\it X-ray study of the structure of phospholipid monolayers on the water surface}, JETP Lett. 106, 534 (2017); arXiv:1712.01607 [cond-mat.soft].

\bibitem{14}
A.~M.~Tikhonov, {\it Compact layer of alkali ions at the surface of colloidal silica}, J. Phys. Chem. C 111, 930 (2007); arXiv:cond-mat/0609437 [cond-mat.soft].

\bibitem{15}
A.~M.~Tikhonov, {\it  Ion-size effect at the surface of a silica hydrosol}, J. Chem. Phys. 130, 024512 (2009); arXiv:0808.3940 [cond-mat.other].

\bibitem{16}
V.\,E.\,Asadchikov, V.\,V.\,Volkov, Yu.\,O.\,Volkov, K.\,A.\,Dembo, I.\,V.\,Kozhevnikov, B.\,S.\,Roshchin,
D.\,A.\,Frolov, and A.\,M.\,Tikhonov, {\it Condensation of silica nanoparticles on a phospholipid membrane}, JETP Lett. 94, 585 (2011); arXiv:1111.0955 [cond-mat.soft].

\bibitem{17}
A.~M.~Tikhonov, V.~E.~Asadchikov, Yu.~O.~Volkov, B.~S.~Roshchin, V.~Honkimaki, M.~Blanco, {\it Model-independent X-ray scattering study of a silica sol surface}, JETP Lett. 107, 384 (2018); arXiv:1804.07670 [cond-mat.soft].

\bibitem{18}
G.~Palasantzas, {\it  Roughness spectrum and surface width of self-affine fractal surfaces via the K-correlation model}, Phys. Rev. B 48, 14472 (1993).

\bibitem{19}
A. ~Braslau, P.~S.~Pershan, G.~Swislow, B.~M.~Ocko, and J.~Als-Nielsen, {\it Capillary waves on the
surface of simple liquids measured by X-ray reflectivity}, Phys. Rev. A 38, 2457 (1988).

\bibitem{20}
L.~I.~Goray, V.~E.~Asadchikov, B.~S.~Roshchin, Yu.~O.~Volkov, A.~M.~Tikhonov, {\it First detection
of X-ray whispering gallery modes at the surface meniscus of a rotating liquid}, OSA Continuum 2, 460 (2019).

\bibitem{WG1}
L.~I.~Goray, V.~E.~Asadchikov, B.~S.~Roshchin, Yu.~O.~Volkov, A.~M.~Tikhonov, {\it Analysis of X-ray whispering gallery waves propagating along liquid meniscuses}, Resource-Efficient Technologies 4(1), 47 (2018).

\bibitem{WG4}
L.~I.~Goray, V.~E.~Asadchikov, B.~S.~Roshchin, Yu.~O.~Volkov, A.~M.~Tikhonov, {\it Reflectometry of x-ray whispering gallery waves propagating along liquid meniscuses}, Semiconductors 52(16), 2049 (2018).

\bibitem{21}
A.~M.~Tikhonov, V. E. Asadchikov, Yu. O. Volkov, {\it On the formation of a macroscopically flat phospholipid membrane on a hydrosol
substrate}, JETP Lett. 102, 478 (2015); arXiv:1511.03543 [cond-mat.soft].

\bibitem{22}
A.~M.~Tikhonov,  V.~E.~Asadchikov, Yu.~O.~Volkov, B.~S.~Roshchin, I.~S.~Monakhov, I.~S.~Smirnov, {\it Kinetics of the formation of a phospholipid multilayer on a silica sol surface}, JETP Lett. 104, 873 (2016); arXiv:1702.02368 [physics.bio-ph].

\bibitem{23} 
S.~A.~Pandit and M.~L.~Berkowitz, {\it Molecular dynamics simulation of dipalmitoylphosphatidylserine bilayer with Na$^+$ counterions}, Biophys. J. 82, 1818 (2002).

\bibitem{24}
Yu.~A.~Ermakov, K.~Kamaraju, K.~Sengupta and S.~Sukharev, {\it Gadolinium ions block mechanosensitive channels by altering the
packing and lateral pressure of anionic lipids}, Biophys. J. 98, 1018 (2010).

\bibitem{25}
A.~M.~Tikhonov, V.~E.~Asadchikov, Yu.~O.~Volkov, B.~S.~Roshchin, Yu.~A.~Ermakov, {\it X-ray
reflectometry of DMPS monolayers on a water substrate}, JETP 125, 1051 (2017); arXiv:1712.08488 [cond-mat.soft].

\bibitem{26}
Yu.\,A.\,Ermakov, V.\,E.\,Asadchikov, B.\,S.\,Roschin, Yu.\,O.\,Volkov, D.\,A.\,Khomich, A.\,M.\,Nesterenko, A.\,M.\,Tikhonov, 
{\it Comprehensive study of the Liquid Expanded-Liquid Condensed phase transition in 1,2-dimyristoyl-sn-glycero-3-phospho-L-serine monolayers: surface pressure, Volta potential, X-ray reflectivity, and molecular dynamics modeling}, Langmuir 35(38), 12326 (2019).

\end{thebibliography}
\end{document}